\documentclass[12pt]{article}

\usepackage{multirow}

\usepackage{dcolumn}
\newcolumntype{.}[1]{D{.}{.}{#1}}
\newcolumntype{,}[1]{D{.}{.}{#1}}

\pdfoutput=1
\usepackage[latin1]{inputenc}

\usepackage{a4,amsmath}
\usepackage{amsmath}
\usepackage{amsfonts}
\usepackage{natbib}
\usepackage{amssymb}
\usepackage[english]{babel}
\usepackage{graphicx}
\usepackage{psfrag}
\usepackage{amsbsy}
\usepackage{xcolor}
\usepackage{amsmath,amssymb}
\usepackage{authblk}
\usepackage{amsmath}

\numberwithin{equation}{section}

\newcounter{example}[section]
\def\theesempio{\thesection.\arabic{example}}

\begin{document}
\bibliographystyle{plain}
\pagenumbering{arabic}


\title{\bf{An alternative marginal likelihood estimator for phylogenetic models}}

\author[1]{\bf Serena Arima \footnote{Dipartimento di studi geoeconomici, linguistici,  
statistici e storici per l'analisi regionale, Sapienza Universit\`a di Roma, via del Castro Laurenziano 9, 00161 Roma, \\E-mail:
serena.arima@uniroma1.it}}
\author[2]{\bf Luca Tardella}

\affil[1]{Dipartimento di studi geoeconomici, linguistici,  
statistici e storici per l'analisi regionale, Sapienza Universit\`a di Roma

via del Castro Laurenziano 9, 00161 Roma,

E-mail: serena.arima@uniroma1.it}
\affil[2]{Dipartimento di Statistica, Probabilit\`a e Statistiche
Applicate, Sapienza Universit\`a di Roma

p.le Aldo Moro, 5, 00161 Roma,

E-mail: luca.tardella@uniroma1.it} 

\maketitle

\begin{abstract}
Bayesian phylogenetic methods are generating 
noticeable enthusiasm in the field of molecular
systematics. Many phylogenetic models are often 
at stake and different approaches are used
to compare them  within a Bayesian framework. 
The Bayes factor, defined as the ratio of the marginal likelihoods of
two competing models, plays a key role in Bayesian model selection.
We focus on an alternative estimator of the 
marginal likelihood 
whose computation is still a challenging problem. 
Several computational solutions have been proposed none 
of which can be considered 
outperforming the others 
simultaneously 
in terms 
of simplicity of implementation, 
computational burden 
and precision of the estimates. 
Practitioners and researchers, often led by available software,
have privileged so far  
the simplicity of the harmonic mean estimator 
(HM) and the arithmetic mean estimator (AM). 
However it is known that 
the resulting estimates of the 
Bayesian evidence in favor of one model are 
biased and 
often 
inaccurate
up to having an infinite variance
so that 
the reliability of 
the corresponding conclusions is doubtful. 
Our new implementation of the 
generalized harmonic mean 
(GHM) idea
recycles MCMC simulations from the posterior,
shares the computational simplicity of the 
original HM estimator, 
but, unlike it, overcomes
the infinite variance issue.
The alternative estimator
is 
applied to 
simulated phylogenetic data
and 
produces fully satisfactory results 
outperforming those simple estimators currently 
provided by most of the publicly available software. 
\end{abstract}

\textit{keywords}: Bayes factor, harmonic mean, importance sampling, marginal likelihood, phylogenetic models.

\section{Introduction}

The theory of evolution states that all organisms are related through 
a history of common ancestor
and that life on Earth diversified in a tree-like pattern connecting all living species.
Phylogenetics aims at inferring the tree that better represents the evolutionary relationships
among species studying differences and similarities in their genomic sequences.
Alternative tree estimation methods 
such as parsimony methods (\cite{Felsenstein:2004}, chapter 7) and distance methods 
\citep{Fitch:1967, Cavalli-Sforza:1967} have been proposed. In this paper, we will focus on 
stochastic models for substitution rates and we address
the model choice issue within a fully Bayesian framework proposing
an alternative 
model evidence
estimation procedure. The paper is organized as follows: in Section \ref{sec:01}
we briefly review basic phylogenetic concepts and the Bayesian inference for substitution models.  In Section
\ref{sec:02} we focus on the model selection issue for substitution models and discuss
some 
 available computational tools for Bayesian model evidence. 
One of the most popular tool for computing model evidence 
in phylogenetics
is the Harmonic Mean (HM) estimator proposed by \cite{Newton:1994}
as an easy-to-apply instance of a more general class 
of estimators called Generalized Harmonic Mean (GHM). 
An  alternative version of GHM is considered 
in Section \ref{sec:03}.
It has been introduced in 
\cite{Petris:2007} under the name of 
Inflated Density Ratio (IDR)
and its implementation for substitution models 
is described in Section \ref{sec:04}.
Numerical examples and comparative performance are given in Section \ref{sec:05}. 
 We conclude with a brief discussion in Section \ref{sec:07}.

\section{Substitution models: a brief overview}
\label{sec:01}

Phylogenetic data consists of homologous DNA strands or protein sequences of related species. 
Observed data consists of a nucleotide matrix $X$ with $n$ rows representing species 
and $k$ columns representing sites.
Comparing 
DNA sequences of
two 
related 
species, we define \textit{substitution} the replacement 
in the same situs
of one nucleotide in one species by another one in the other species. 
The stochastic
models describing this replacement process are called 
\textit{substitution models}.
A phylogeny or a phylogenetic tree is a representation of the genealogical relationships among species, 
also called \textit{taxa} or \textit{taxonomies}. 
Tips (leaves or external
nodes) represent the present-day species, 
while internal nodes usually represent
extinct ancestors for which genomic sequences are 
no longer
available. The ancestor of all
sequences is the root of the tree. 
The branching pattern of a tree is called \textit{topology}, 
and is denoted with $\tau$, 
while the 
lengths $\nu_\tau$
of the branches of the tree $\tau$
represent
the time periods 
elapsed until a new substitution occurs.\\
DNA substitution models are probabilistic models which aim at modeling changes
between nucleotides in homologous DNA strands. 
Changes at each site occur at random times.
Nucleotides at different sites 
are usually assumed to evolve independently each
other. 
For a fixed site, nucleotide replacements
over time 
are modeled by a 4-state Markov process, in which each state represents
a nucleotide. 
The 
Markov process indexed with time $t$ 
is completely specified by a substitution rate matrix $Q(t)={r_{ij}(t)}$:
each element $r_{ij}(t)$, $i \neq j$, represents the
instantaneous rate of substitution from nucleotide $i$ to nucleotide
$j$. The diagonal elements of the rate matrix are defined as
$r_{ii}(t)=\sum_{j \neq i}r_{ij}(t)$
so that $\sum_{j=1}^{4}r_{ij}(t)=0$,
$\forall i$. The 
transition
probability matrix 
$P(t)=\{p_{ij}(t)\}$,
defines the probability of 
changing
from state $i$ to state $j$.
The substitution process is assumed homogeneous over time
so that
$Q(t)=Q$ and $P(t)=P$. 
 It is also 
commonly 
assumed that the substitution process 
at each site 
is
stationary with equilibrium distribution $\Pi =(\pi_{A},\pi_{C},\pi_{G},\pi_{T})$ and
time-reversible, that is
\begin{equation}
\centering \pi_{i}r_{ij}=\pi_{j}r_{ji} \label{balance}
\end{equation}
where $\pi_{i}$ is the proportion of time the Markov chain spends in
state $i$ and $\pi_{i}r_{ij}$ is the amount of flow from state $i$
to $j$. Equation (\ref{balance}) is known as \textit{detailed-balance
condition} and means that flow between any two states in the
opposite direction is the same. 
Following the notation in \cite{Hudelot:2003}, we define
$r_{ij}(t)=r_{ij}=\rho_{ij}\pi_{i}$, $\forall i \neq j$, where $\rho_{ij}$ is
the transition rate from nucleotide $i$ to nucleotide $j$.
This reparameterization is particularly useful for the specification
of substitution models, since it makes clear the distinction between the nucleotide frequencies
$\pi_{A},\pi_{G},\pi_{C},\pi_{T}$ and substitution rates
$\rho_{ij}$, allowing 
to spell out 
different assumptions on evolutionary patterns.
The most general time-reversible
nucleotide substitution model is the so-called \textbf{GTR} 
defined 
 by the following rate matrix
\begin{eqnarray}
\centering
 \mbox{$Q$}&=&\left(\begin{array}{c c c c}
$-$ & \rho_{AC}\pi_{C} & \rho_{AG}\pi_{G} & \rho_{AT}\pi_{T}\\
\rho_{AC}\pi_{A} &$-$ & \rho_{CG}\pi_{G} & \rho_{CT}\pi_{T}\\
\rho_{AG}\pi_{A} & \rho_{CG}\pi_{C} & $-$ & \rho_{GT}\pi_{T}\\
\rho_{AT}\pi_{A} & \rho_{CT}\pi_{C} & \rho_{GT}\pi_{G} & $-$
\end{array} \right)
\label{Substitution-Matrix Q-1}
\end{eqnarray}
and more thoroughly illustrated  
in \cite{Lanave:1984}. 
Several substitution models can be obtained
simplifying the $Q$ matrix reflecting specific biological assumptions: 
the simplest one is the \textbf{JC69} model, originally proposed in \cite{Jukes:1969},
which assumes that all nucleotides are interchangeable and have the same rate of change, 
that is $\rho_{ij}=\rho \    \ \forall i, j$ and $\pi_{A}=\pi_{C}=\pi_{G}=\pi_{T}$.\\
In this paper, 
for illustrative purposes 
we will consider 
only instances of 
{GTR} and {JC69} models.
One can look at 
\cite{Felsenstein:2004} and \cite{Yang:2006}
for a wider 
range 
of 
alternative 
substitution models.

\subsection{Bayesian inference for substitution models}

The parameter space of a phylogenetic model 
can be represented as
$$\Omega=\{\tau, \nu_{\tau}, \theta\}$$ 
where
$\tau \in{\cal T}$ is the tree topology, 
$\nu_{\tau}$ the set of branch lengths of topology $\tau$,
and $\theta=(\rho, \pi)$ the parameters of the rate matrix.
We denote $N_{T}$ 
the cardinality of $\cal T$.
Notice that $N_{T}$ is a huge number even for few species. 
For instance 
with $n=10$ species
there are about $N_{T} \approx 2 \cdot 10^{6}$
different trees.\\
Observed data consists of a nucleotide matrix 
$X$
once specified the substitution model $M$,
the likelihood $p(X|\tau, \nu_{\tau}, \theta, M)$ 
can be computed 
using the pruning algorithm, a practical and efficient recursive algorithm 
proposed in \cite{Felsenstein:1981}.
One can then make inference on the unknown model parameters
looking for the 
values  which 
maximize the likelihood. 
Alternatively one can adopt
a Bayesian approach 
where the parameter space is endowed with 
a joint prior distribution 
$\pi(\tau, \nu_{\tau},\theta)$ 
on the unknowkn parameters 
and 
the likelihood is 
used to 
update the prior uncertainty
about 
$(\tau, \nu_{\tau},\theta)$ 
following
the Bayes' rule:
$$p(\theta,\nu_{\tau},\tau |X,M)= \frac{p(X|\tau, \nu_{\tau}, \theta, M)\pi(\tau, \nu_{\tau},\theta)}{m(X|M)}$$
where
$$m(X|M)=\sum_{\tau \in{\cal T}}\int_{\nu_{\tau}}\int_{\theta}p(X|\tau,\nu_{\tau},\theta, M) \pi(\tau, \nu_{\tau},\theta)d\nu_{\tau}d\theta$$
The resulting distribution $p(\tau, \nu_{\tau}, \theta|X,M)$ 
is the posterior distribution 
which coherently combines 
prior believes and data information. 
Prior believes are 
usually
conveyed 
as 
$\pi(\tau, \nu_{\tau},\theta)= \pi(\tau)\pi(\nu_{\tau}) \pi(\theta)$.
The denominator of the Bayes' rule
$m(X|M)$
is the marginal likelihood of model $M$ and
it plays a key role in 
 discriminating among alternative models.

More precisely, suppose, we aim at 
comparing two competing substitution models $M_{0}$ and $M_{1}$. 
The Bayes Factor is defined as the ratio of the marginal likelihoods
as follows 
$$BF_{10}=\frac{m(X|M_{1})}{m(X|M_{0})}$$
where, for $i=0,1$
\begin{equation}
c^{(i)}=m(X|M_{i})=\sum_{\tau \in {\cal {T}}}\int_{\nu^{(i)}_{\tau}}\int _{\theta^{(i)}} p(X|\theta,\tau,M_{i})\pi(\theta^{(i)})d\theta^{(i)} \pi(\nu_{\tau}^{(i)}|\tau) 
d \nu_{\tau}^{(i)} \pi(\tau)
\label{normcost:phylo}
\end{equation}
Numerical guidelines for interpreting 
the evidence scale are given in \cite{Kass:1995}.
Values of $BF_{10}>1$ ($log(BF_{10})>0)$ 
can 
be 
considered
as 
evidence in favor of $M_{1}$ 
but only a value of $BF_{10}>100$ ($log(BF_{10}>4.6$) can 
be really considered as decisive.

Most of the times 
the posterior distributions $p(\tau, \nu_{\tau}, \theta|X,M_i)$ 
and marginal likelihoods 
are 
not analytically computable
but can be approached through appropriate approximations.
Indeed, over the last ten years,
powerful numerical methods based on Markov Chain Monte Carlo (MCMC)
have been developed, allowing one to 
carry out 
Bayesian 
inference under a large category of probabilistic models, 
even when dimension of the parameter space is very large.
Indeed, several ad-hoc  MCMC algorithms have been 
tailored 
for phylogenetic models 
\citep{Larget:1999, Li:2000} and
are currently implemented in publicly available software such as 
in \textsc{MrBayes} \citep{Ronquist:2003}  and \textsc{PHASE}
\citep{phase:2006}.\\

\section{Model selection for substitution models} 
\label{sec:02}

Given the variety of possible stochastic substitution 
mechanisms, an
important issue of any model-based phylogenetic analysis is to
select the model which is most supported by the data. 
Several model selection procedures
have been proposed depending also on the inferential approach. 
A 
classical 
approach to model selection 
for choosing between alternative nested models 
is to perform the hierarchical likelihood ratio test (LRT)
\citep{Posada:2001}. 
A number of
popular programs allow users to compare pairs of models using this
test such as PAUP \citep{Swafford:2003}, PAML \citep{Yang:2007} and the
\textsf{R} package \textsf{APE} \citep{R:2008}. However, 
\cite{Posada:2004} have shown some drawbacks of performing systematic LRT
for model selection in phylogenetics. This is because
the model that is finally selected can depend on the order in which
the pairwise comparisons are performed \citep{Pol:2004}. 
Moreover, it is well-known that LRT tends to favor parameter rich models. \\
The Akaike Information Criterion (AIC) is another model-selection
criterion commonly used also in phylogenetics \citep{Posada:2004}:
one of the advantages of the AIC is that it allows 
to compare nested as well as 
non nested models
and it can be easily implemented. 
However, 
also 
the AIC
tends to favor parameter-rich models.
To overcome this
selection bias
one can use 
the Bayesian Information Criterion (BIC)
\citep{Schwartz:1978}
which 
better
penalizes parameter-rich
models.  

Sometimes
these criteria 
applied to the same data can end up selecting 
very different substitution models, as shown in
\cite{Abdo:2005}. Indeed  they compare
ratios of likelihood values penalized for an increase in the
dimension of one of the models, without directly 
accounting 
for
uncertainty in the estimates of model parameters.
The latter aspect is
addressed 
within a fully Bayesian framework
through the use of the Bayes Factor. 
Bayes Factor 
directly incorporates this uncertainty and 
its meaning 
is more intuitive than
other methods since it 
can be directly used 
to assess
the comparative evidence provided by the data 
in terms of the 
most probable model 
under equal prior model probabilities. \\
Bayes Factor for comparing phylogenetic models was
first introduced in \cite{Sinsheimer:1996} and 
\cite{Suchard:2001}. Since then its popularity 
in phylogenetics 
has grown 
so that 
some publicly available software 
provide in their standard output
approximations of 
marginal likelihoods for 
model evidence and 
Bayes Factor evaluation.
Indeed the complexity of 
phylogenetic models
and the computational 
burden in the light of
high-dimensional
parameter space 
make the problem of finding 
alternative and more efficient computational 
strategy
for computing Bayes Factor
still open and in continuous development  \citep{Lartillot:2007, Ronquist:2010}.\\

\subsection{Available
computational tools for Bayesian model evidence}

The computation of the marginal likelihood
$m(X|M)$
of a
phylogenetic model
$M$
is not straightforward.
It 
involves 
integrating 
the likelihood 
over 
$k$-dimensional subspaces for the branch length parameters 
$\nu_{\tau}$ and the substitution rate matrix $\theta=(\rho, \pi)$
and eventually summing over all possible topologies.\\
Most of the marginal likelihood estimation methods 
proposed in the literature have been applied extensively also in
molecular phylogenetics \citep{Minin:2003,Lartillot:2007,Suchard:2001}. 
Among these methods, many of them are valid
only under very specific conditions. For instance, the Dickey-Savage
ratio \citep{Verdinelli:1995} applied in phylogenetics in
\cite{Suchard:2001}, assumes nested models. Laplace approximation
\citep{Kass:1995} and BIC
\citep{Schwartz:1978}, applied in phylogenetics firstly in
\cite{Minin:2003}, require large sample approximations around the
maximum likelihood, which can be sometimes 
difficult to compute or approximate for very complex models. 
A recent appealing variation of the Laplace 
approximation has been proposed in \cite{Rodrigue:2007}: however, 
its 
applicability  and performance
are endangered 
when the posterior distribution 
deviates  from normality 
and the maximization of the likelihood can be neither straightforward
nor accurate.\\
The reversible jump approach \citep{Green:1995, Bartolucci:2006}
is another MCMC 
option
applied to phylogenetic
model selection in \cite{Huelsenbeck:2004}.
Unfortunately 
the implementation of this algorithm
is not straightforward for the end user and 
it often requires appropriate delicate tuning
of the Metropolis Hastings proposal.
Moreover its implementation
suffers extra difficulties
when comparing models
based on an entirely different parametric rationale \citep{Lartillot:2007}.\\
As recently pointed out in \cite{Ronquist:2010}
among 
the most widely used methods for estimating the marginal likelihood
of phylogenetic models are the thermodynamic integration, 
also known as path sampling, 
and the harmonic mean approach. 
The thermodynamic integration 
reviewed in \cite{Gelm:Meng:simu:1998} and first applied in a phylogenetic context 
in \cite{Lartillot:2006} produces reliable
estimates of Bayes Factors of phylogenetic models 
in a large varieties of models.
Although this method has the advantage of general applicability,
it can incur high computational costs and 
may require specific adjustments. 
For certain model comparisons, a full thermodynamic integration
may take weeks on a modern desktop computer,
even under a fixed tree topology for small single
protein data sets \citep{Rodrigue:2007, Ronquist:2010}. 
On the other hand, the HM estimator
can be easily computed and it does not 
demand further computational efforts 
other than those already 
made to draw inference on 
model parameters, since it 
only needs simulations from the posterior distributions.
However, it is well known that the HM estimator
is unstable since it can end up with an infinite variance.
As highlighted by \cite{Ronquist:2010}, 
thermodynamic integration and reversible jump are, until now, the
most accurate tools for computing the marginal likelihood. 
However,
until these methods become more user-friendly and
more widely available, 
simple tools for exploring in a quicker way the more interesting models
are 
useful. 
For this reason in the next sections we focus on 
an alternative generalized harmonic mean estimator, the IDR estimator, which shares the computational
simplicity of HM estimator but, unlike it, 
better 
copes  with
the infinite variance issue.
Its simple implementation makes the IDR estimator a useful and 
more 
reliable method 
for 
easily 
comparing competing substitution models. 
It can be used 
also 
as a confirmatory tool
even in those models 
for which more complex estimation methods, such as the path sampling, can be applied.

\subsection{Harmonic Mean estimators}

We introduce the 
basic ideas and formulas for the 
class of estimators known as 
Generalized Harmonic Mean (GHM).
Since the marginal likelihood is nothing but the
normalizing constant of the unnormalized posterior density,
we illustrate 
the GHM estimator as a general solution for estimating 
the normalizing constant 
of a non-negative, integrable density $g$  defined as
\begin{eqnarray}
\label{ccc}
c&=&\int_{\Omega}g(\theta)d\theta
\end{eqnarray}
where $\theta \in \Omega \subset \Re^k$ 
and $g(\theta)$ is the unnormalized version of the 
probability distribution $\tilde{g}(\theta)$.
The GHM estimator of 
$c$ is based on the following identity
\begin{equation}
c=
\frac{1}{E_{\tilde{g}}\left[\left(\frac{g(\theta)}{f(\theta)}\right)^{-1}\right]}
\label{HM-estimator}
\end{equation}
where $f$ is a convenient instrumental Lebesgue integrable function 
which is only 
required to have a support which is contained 
in that of $g$ and to satisfy 
\begin{eqnarray}
\label{ghmconstraint}
\int_{\Omega}f(\theta)d\theta=1.
\end{eqnarray}
The GHM estimator, denoted as $\hat{c}_{GHM}$
is the empirical counterpart of 
(\ref{HM-estimator}), namely
\begin{equation}
\label{GHM}
 \hat{c}_{GHM}=
\frac{1}{\frac{1}{T}\sum_{t=1}^{T}\frac{f(\theta_{t})}{g(\theta_{t})}}.
\end{equation}
where $\theta_{1}, \theta_{2}, ..., \theta_{T}$ are sampled from $\tilde{g}$.
In Bayesian inference the very first instance of such 
GHM estimator was introduced 
in \cite{Gelf:Dey:baye:1994}
to estimate the 
marginal likelihood considered as the 
normalizing constant 
of the unnormalized posterior density 
$g(\theta)=\pi(\theta) L(\theta)$.
Hence,
taking $f(\theta)=\pi(\theta)$ one obtains as
special case of (\ref{GHM}) the 
Harmonic Mean estimator
\begin{equation}
\label{HM}
 \hat{c}_{HM}=
\frac{1}{\frac{1}{T}\sum_{t=1}^{T}\frac{1}{L(\theta_{t})}}
\end{equation}
which can be
easily computed by recycling simulations 
$\theta_1,...,\theta_T$
from the target posterior 
distribution $\tilde{g}(\theta)$
available from MC or MCMC sampling scheme.
This probably explains the original 
enthusiasm in favor of 
$ \hat{c}_{HM}$ which indeed was considered 
a potential convenient competitor 
of the standard Monte Carlo 
Importance Sampling 
estimate given by the  (Prior) Arithmetic Mean (AM) estimator 
\begin{equation}
\label{AM}
 \hat{c}_{AM}=
\frac{1}{T} \sum_{t=1}^{T} {L(\theta_{t})}
\end{equation}
where 
$\theta_1,...,\theta_T$ are sampled from the prior $\pi$.

The implementation of 
\label{HM} and ,more generally,
\eqref{GHM}
requires a relatively 
light computational burden 
hence 
reducing computing time 
with respect to thermodynamic integration.
The simplicity of the computation
has then favored the widespread use of the Harmonic Mean estimator
with respect to more complex methods.
In fact, the Harmonic Mean estimator is implemented in
several Bayesian phylogenetic software as shown in Table~\ref{Tab1} and
recent biological papers \citep{Yamanoue:2008,Wang:2009,Norman:2009} report
the HM as a routinely used model selection tool. 

\begin{center}
\textbf{Table 1 about here}
\end{center}

However, both 
$\hat{c}_{AM}$ and 
$\hat{c}_{HM}$ 
can end up with a very large
variance and 
unstable behavior.
This fact cannot be considered as 
an unusual
exception
but it often occurs and the reason 
for that 
can be argued
on a theoretical ground. 

For $\hat{c}_{AM}$ the erratic behavior 
is simply explained by the fact that 
the likelihood 
usually 
gives support to a region with low 
prior weight hence sampling from the prior yields 
low chance to hit high likelihood region and large chance 
to hit much lower likelihood region ending up in a large variance
of the estimate $\hat{c}_{AM}$.
Indeed,
starting from the 
original paper 
\cite{Newton:1994}, 
(see in particular R. Neal's discussion)
it has been shown 
that
even in very simple and standard gaussian models 
also the 
HM estimator can end up having an infinite variance
hence yielding 
unreliable approximations. 
This fact raises sometimes the question 
whether they are  reliable tools and 
certainly 
has encouraged researchers 
to look for alternative solutions.
Several generalizations and
improved alternatives have been 
proposed and recently reviewed in 
\cite{Raftery:2007}.

In the following sections 
we will consider a new marginal 
likelihood estimator, 
the Inflated Density Ratio (IDR) estimator, proposed in 
\cite{Petris:2007}, 
which is a particular instance of the
Generalized Harmonic Mean (GHM) approach.
This new estimator basically shares
the original simplicity and the computation feasibility 
of the HM estimator but, unlike it, it
can guarantee
important theoretical 
properties, such as 
a bounded variance.

\section{IDR: Inflated Density Ratio estimator}
\label{sec:03}

\begin{center}
\begin{figure}[htbp]
\includegraphics[scale=0.4]{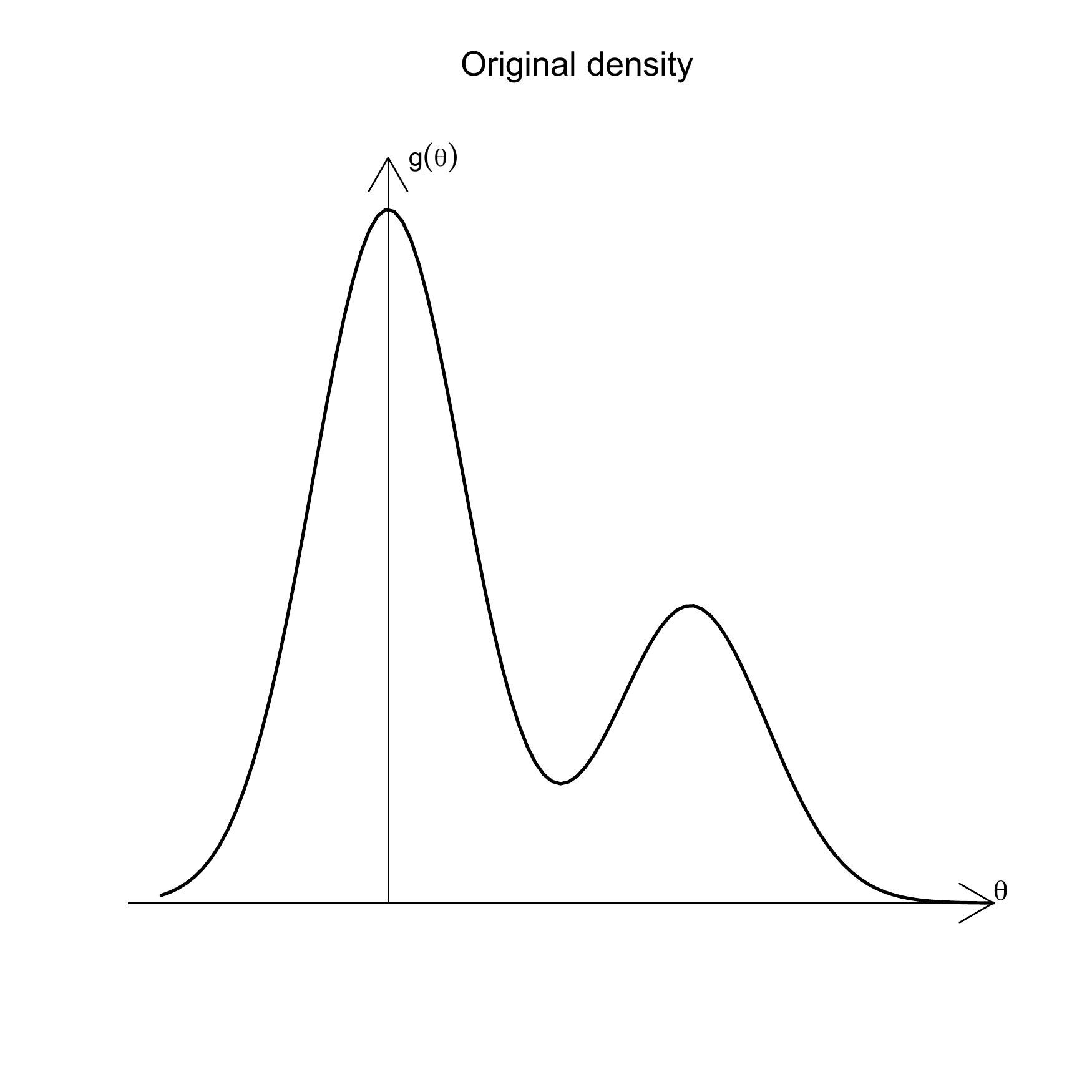}
\quad \hspace{-0.5cm}
\includegraphics[scale=0.4]{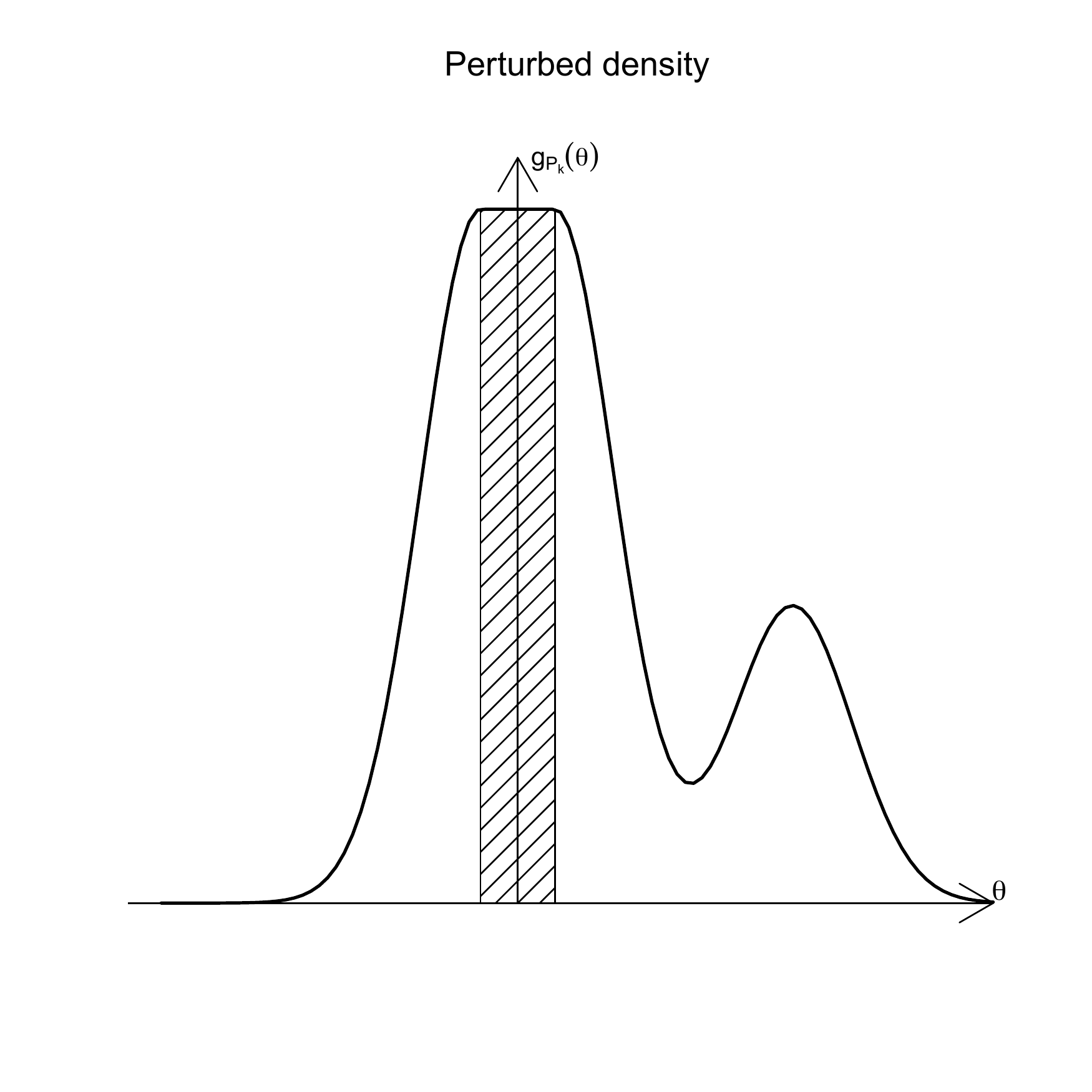}
\caption{\em Left Panel: original density $g$ with total mass $c$. Right Panel:
perturbed density $g_{P_{k}}$ defined as
in (\ref{gpk}).
The total mass of the perturbed density is then $c+k$. The shaded 
area correspond to the inflated mass $k$ with $k= 2\cdot r_k\cdot g(0)$ as in 
(\ref{gpk}).}
\label{Fig.1}
\end{figure}
\end{center}

The inflated density ratio estimator 
is a different formulation of the
GHM estimator, 
based on a particular choice of
the instrumental density 
$f(\theta)$
as 
originally proposed in \cite{Petris:2007}.
The instrumental $f(\theta)$
is obtained through  a
perturbation of the original target function
$g$. The 
perturbed density,
denoted with $g_{P_{k}}$,
is defined 
so that its total mass has some
known functional relation to the total mass $c$ of the target
density $g$ as in \eqref{ccc}. 
In particular, $g_{P_{k}}$ is obtained as a parametric inflation of
$g$ so that
\begin{center}
\begin{equation}
\int_{\Omega}g_{P_{k}}(\theta)=c+k
\end{equation}
\label{g_P_{k}}
\end{center}
where $k$ is a known inflation constant which can be arbitrarily
fixed. The perturbation 
device 
comes from an original idea in 
\cite{Petris:2003}
and is deteiled in \cite{Petris:2007}
for unidimensional and multidimensional densities. 
In the unidimensional case 
the perturbed density is 
\begin{equation}
\label{gpk}
g_{P_{k}}(\theta) = \left\{
\begin{array}{ll}
g(\theta+r_{k}) & \mbox{if $\theta < -r_{k}$}\\
g(0)& \mbox{if $-r_{k}\leq \theta \leq r_{k}$}\\
g(\theta-r_{k}) & \mbox{if $\theta > r_{k}$}
\end{array}
\right.
\end{equation}
with  $2r_{k}=\frac{k}{g(0)}$ 
corresponding to 
the length of the interval
centered around the origin
where the density is kept constant. 
In Figure \ref{Fig.1} 
one can 
visualize 
how the perturbation acts. 
The perturbed density allows one to define 
an instrumental density 
$f_{k}(\theta)=\frac{g_{P_{k}}(\theta)-g(\theta)}{k}$ 
which satisfies the requirement 
\eqref{ghmconstraint}
needed to define the 
GHM estimator as in 
\eqref{GHM}.
The Inflated Density Ratio estimator $\hat{c}_{IDR}$ for $c$ 
is then 
obtained 
as follows
\begin{center}
\begin{equation}
\hat{c}_{IDR}=\frac{k}{\frac{1}{T}\sum_{t=1}^{T}\frac{g_{P_{k}}(\theta_{t})}{g(\theta_{t})}-1}
\end{equation}
\label{c_ID}
\end{center}
where $\theta_{1},...,\theta_{T}$ is a sample from the normalized
target density $\tilde{g}$. The use of the perturbed density 
as importance function
leads to some advantages with respect
to the other instances of $c_{GHM}$
proposed in the literature. 
In fact 
$\hat{c}_{IDR}$ defined as in 
(\ref{c_ID}) yields a finite-variance estimator 
under mild sufficient conditions and 
a wide range of $g$ densities
\cite{Petris:2007} (Lemma 1, 2 and 3).
Notice that in order for the perturbed density 
$g_{P_{k}}$
to be defined 
it is required that the original density $g$ 
has full support in $\Re^k$.
Moreover, the use of a parametric perturbation 
makes the method more flexible and efficient 
with a moderate extra computational effort.

Like all methods based on importance sampling strategies, the
properties of the estimator $\hat{c}_{IDR}$ strongly depend on the
ratio $\frac{g_{P_{k}}(\theta)}{g(\theta)}$. To 
evaluate its performance one can 
use
an asymptotic  approximation (via standard delta-method) 
of the Relative Mean Square Error of the estimator
\begin{center}
\begin{equation}
RMSE_{\hat{c}_{IDR}}=\sqrt{E_{\tilde{g}}\left[\left(\frac{\hat{c}_{IDR}-c}{c}\right)^{2}\right]}\approx \frac{c}{k}\sqrt{Var\left[\frac{g_{P_{k}}(\theta)}{g(\theta)}\right]} = 
RMSE_{\hat{c}_{IDR, Delta}}
\end{equation}
\label{RMSE}
\end{center}
which can be ultimately estimated
as follows:
\begin{center}
\begin{equation}
 \widehat{RMSE}_{\hat{c}_{IDR, Delta}} = \frac{\hat{c}_{IDR}}{k}\sqrt{\widehat{Var}_{\tilde{g}}\left[\frac{g_{P_{k}}(\theta)}{g(\theta)}\right]}
\label{RMSE_hat}
\end{equation}
\end{center}
where $\widehat{Var}_{\tilde{g}}$ is the 
observed sample variance of the ratio
${g_{P_{k}}(\theta)}/{g(\theta)}$.\\
The expression in Equation (\ref{RMSE_hat}) clarifies the key role
of the choice of $k$ with respect to the error of the estimator: 
for $k\rightarrow 0$, the variance of the ratio
$\frac{g_{P_{k}}(\theta)}{g(\theta)}$ tends to 0, since $g_{P_{k}}$
is very close to $g$, but $\frac{c}{k}$ tends to infinity: in other
words, if
$Var_{\tilde{g}}\left[\frac{g_{P_{k}}(\theta)}{g(\theta)}\right]$
would favor as little values of $k$ as possible, 
$\frac{1}{k}$ acts in the opposite direction. 
In order to address the choice of $k$, \cite{Petris:2007} suggested
to choose the perturbation $k$ which minimizes the estimation error
\eqref{RMSE_hat}.
In practice, 
one can calculate the values 
of the estimator for a grid of different perturbation values,
$\hat{c}_{IDR}(k)$, $k=1,...,K$ and choose the optimal $k^{opt}$ as
the $k$ for which $\widehat{RMSE}_{\hat{c}_{IDR}(k)}$ is minimum.
This procedure for the calibration of $k$ requires 
iterative evaluation of $\hat{c}_{IDR}(k)$
hence is relatively heavier than 
the HM estimator, but it does not require extra simulations
which in the phylogenetic context is often the 
main time-consuming part. 
Hence, 
the computational cost is alleviated by the fact that 
one uses the same sample from $\tilde{g}$ 
and the only quantity to
be evaluated $K$ times is the inflated density $g_{P_{k}}$. 
Once obtained the ratio
of the perturbed and the original density, the computation of 
$\hat{c}_{IDR}(k)$ is
straightforward. 

For practical purposes, the computation of the
inflated density 
when the support of $g$ is the whole 
$\Re ^k$
can be easily implemented in \textsf{R}
using a function available at 
\verb@http://sites.google.com/site/idrharmonicmean/home@.
In \cite{Petris:2007} and \cite{Arima:2009} it has been shown that 
in order to improve the precision of $\hat{c}_{IDR}$ 
it is recommended to standardize the simulated data $\theta_{1},...,\theta_{T}$
with respect to a (local) mode and the sample variance-covariance matrix
so that the corresponding density has a local mode at the origin and
approximately standard variance-covariance matrix.
This is 
automatically implemented in the 
publicly available \verb@R@ code. 



In order to assess its effectiveness, 
the IDR method
has been applied
to simulated data from differently shaped  distributions
for which the normalizing constant is known. 
As shown in \cite{Petris:2007},
the estimator produces fully convincing results 
with 
simulated data from 
several 
known distributions,
even for a 100-dimensional multivariate 
Gaussian
distribution.
In terms of estimator precision, 
these results are 
comparable with those in \cite{Lartillot:2006} 
obtained with the thermodynamic integration.
In \cite{Arima:2009} 
simple antithetic variate tricks 
allow the IDR 
estimator to perform well even
for those distributions with severe variations from 
the symmetric Gaussian case 
such as asymmetric and even some 
multimodal distributions. 
Table~\ref{Tab2} shows the estimates obtained by applying
the IDR method in several controlled scenarios: 
the method correctly reproduces 
the true value of the normalizing constant for different shape 
and dimension of the target function. 
Some real data implementation with standard generalized linear models 
have been also reported in 
\cite{Petris:2007}.
In the next Section, we extend the IDR method in order to  
use $\hat{c}_{IDR}$ in more complex settings such as phylogenetic models.

\begin{center}
\textbf{Table 2 about here}
\end{center}

\section{Implementing IDR for substitution models with fixed topology}
\label{sec:04}

We extend the
Inflated Density Ratio approach 
in order to
compute the marginal likelihood of
phylogenetic models. 
In this section we show
how to compute the marginal likelihood when 
it involves integration of 
substitution model parameters $\theta$ and the
branch lengths $\nu_{\tau}$ which are both defined in 
continuous subspaces.
Indeed the approach can be used
as a building block to integrate also over 
the tree topology $\tau$.
\\
For a fixed topology $\tau$ and a sequence alignment $X$,
the parameters 
of a phylogenetic model $M_\tau$ 
are denoted as $\omega = (\theta, \nu_{\tau}) \subset \Omega_\tau$.
The joint posterior distribution on $\omega$ is given
by
\begin{equation}
p(\theta,\nu_{\tau}|X,M_\tau)=\frac{p(X|\theta,\nu_{\tau},M_\tau)\pi(\theta)\pi(\nu_{\tau})}{
m(X|M_\tau)}
\label{Post-Prob-FixedTree}
\end{equation}
where
\begin{equation}
m(X|M_{\tau})=\int_{\theta}\int_{\nu_{\tau}}p(X|\theta,\nu_{\tau},M_\tau)\pi(\theta)\pi(\nu_{\tau})d\theta d\nu_{\tau}
\label{Marginal-Likelihood-Fixed}
\end{equation}
is the marginal likelihood we aim at estimating.\\
When 
the topology $\tau$ is fixed,
the parameter space
$\Omega_\tau$ is 
continuous.
Hence, in order to apply the IDR method
we only need the following
two ingredients:
\begin{itemize}
\item[$\bullet$] a sample $(\theta^{(1)},\nu_{\tau}^{(1)}),...,(\theta^{(T)},\nu_{\tau}^{(T)})$ from the posterior distribution,
$p(\theta,\nu_{\tau}|X,M_{\tau})$
\item[$\bullet$] the likelihood and the prior distribution evaluated at each posterior
sampled value $(\theta^{(k)},\nu_{\tau}^{(k)})$, that is
$p(X|\theta^{(k)},\nu_{\tau}^{(k)},M_\tau)$ and
$\pi(\theta^{(k)},\nu_{\tau}^{(k)})=
\pi(\theta^{(k)})\pi(\nu_{\tau}^{(k)})$
\end{itemize}
The first ingredient is just 
the usual 
output of the Monte 
Carlo Markov Chain 
simulations derived from  model $M$ and 
data $X$. 
The computation of the likelihood 
and the joint prior
is usually already coded within 
available software. 
The first one is accomplished 
through the pruning algorithm 
while computing the prior is 
straightforward.
Indeed a 
 necessary condition for 
the inflation idea 
to 
be implemented 
as prescribed in \cite{Petris:2007}
is that 
the posterior density  
must have
full support on the whole real $k$-dimensional space.
In our phylogenetic models this is not always the case 
hence
we explain 
simple 
and 
fully automatic 
remedies to overcome this kind of obstacle.\\
We start with 
branch length parameters
which 
are constrained to lie in 
the positive half-line.
In that case 
the remedy is straightforward.
One  can reparameterize 
with a simple logarithmic 
transformation 
\begin{equation}
\centering \nu^{'}_{\tau}=log(\nu_{\tau}) \label{Branch-Length}
\end{equation}
so that 
the 
support corresponding to the reparameterized 
density 
becomes
unconstrained.
Obviously the $log(\nu_{\tau})$ reparameterization calls for the 
appropriate 
Jacobian when 
evaluating the corresponding transformed density. 
For model parameters 
with linear constraints like 
the substitution 
$\theta=\{\rho, \pi\}$, a little less obvious
transformation is needed.
In this case 
$\theta=\{\rho, \pi\}$
are subject to the following set of constraints: 
\begin{eqnarray*}
\sum_{i \in \{A,T,C,G\}} \pi_{i}&=&1\\
\sum_{j \in \{A,T,C,G\}} \rho_{ij}\pi_{j} &=& 0 \qquad \qquad 
\forall \  \ i\in \{A,T,C,G\}
\end{eqnarray*}
Similarly to the first simplex constraint 
the last set of constraints together with the reversibility 
can be 
rephrased \citep{shankar:2006} in terms of another simplex constraint 
concerning only the 
extra-diagonal 
entries of the substitution rate matrix 
(\ref{Substitution-Matrix Q-1})
namely 
$$
\rho_{AC}+\rho_{AG}+\rho_{AT}+\rho_{CG}+\rho_{CT}+\rho_{GT} = 1.
$$
In order to bypass 
the constrained nature of the parameter space
we have relied on the so-called
\emph{additive logistic transformation} 
\citep{Tiao:1965, Aitchinson:1986}
which is a 
one-to-one transformation from 
$\mathbb{R}^{D-1}$
to the $(D-1)$-dimensional simplex 
$$S^{D}=\left\{(x_{1},...,x_{D}): x_{1}>0,...,x_{D}>0; x_{1}+...x_{D}=1\right\}.$$ 
Hence we can use its 
inverse, 
called {\em additive log-ratio transformation}, 
which is defined as follows 
$$y_{i}=log\left(\frac{x_{i}}{x_{D}}\right)
\qquad i
=1,...,D-1$$
for any 
$x = (x_{1},..., x_{D}) \in S^{D}$. 
Here 
the $x_i$'s are the $\rho$'s
and 
$D=6$.
Applying these transformations to nucleotide frequencies
$\pi_i$
and to exchangeability parameters $\rho$'s, 
the transformed parameters assume values in the entire real support
and the IDR estimator can be 
applied.
Again the reparameterization calls for the 
appropriate change-of-measure Jacobian when 
evaluating the corresponding transformed density
(see \cite{Aitchinson:1986} for details).
\\

\section{Numerical examples and comparative performance}
\label{sec:05}

In this section the successful implementation 
of the IDR estimator 
is illustrated 
with 
data simulated from
some 
typical phylogenetic models.

Here
IDR method has 
been applied 
using the MCMC output of the 
simulations 
from the posterior distribution obtained using
the \textsf{MrBayes} software;
the likelihood 
has been computed
using
the \textsf{R} package \textsf{PML}
while 
the reparameterization on 
$\Re^k$ 
and 
IDR perturbation $g_{P_{k}}(\theta)$
have called for 
specifically 
developed 
\textsf{R} functions. 
The whole \textsf{R} code
is available upon request from the first author.



Two of
the simplest and most favorite model evidence 
output 
in the publicly available software
are used as benchmarks:
the Harmonic Mean estimator and the Arithmetic Mean estimator.
Indeed, while the former is guaranteed to be a consistent 
estimate of the marginal likelihood,
though possibly with infinite variance,
the latter one 
is consistent only when formula (\ref{AM}) is applied when 
$\theta_1,...,\theta_T$ are sampled from the prior.
Since it is known such a prior AM 
turns out to be 
very unstable and unreliable 
it has 
often 
been replaced 
by a posterior AM where 
$\theta_1,...,\theta_T$ are sampled from the posterior
rather than from the prior. 
In that case 
one must be aware that 
the resulting quantity can be 
interpreted only as a surrogate evidence in favor of one model 
and it should by no means be confused with the rigorous concept of 
marginal likelihood and
related 
to Bayes Factor. 
We now show 
the performance of IDR in two phylogenetic examples. 
Simulated data is used to have a better control of 
what one should expect from 
marginal likelihood and 
the corresponding 
comparative evidence of alternative models.

\subsection{Hadamard 1: marginal likelihood computation}

\begin{center}
\begin{figure}[htbp]
\includegraphics[scale=0.4]{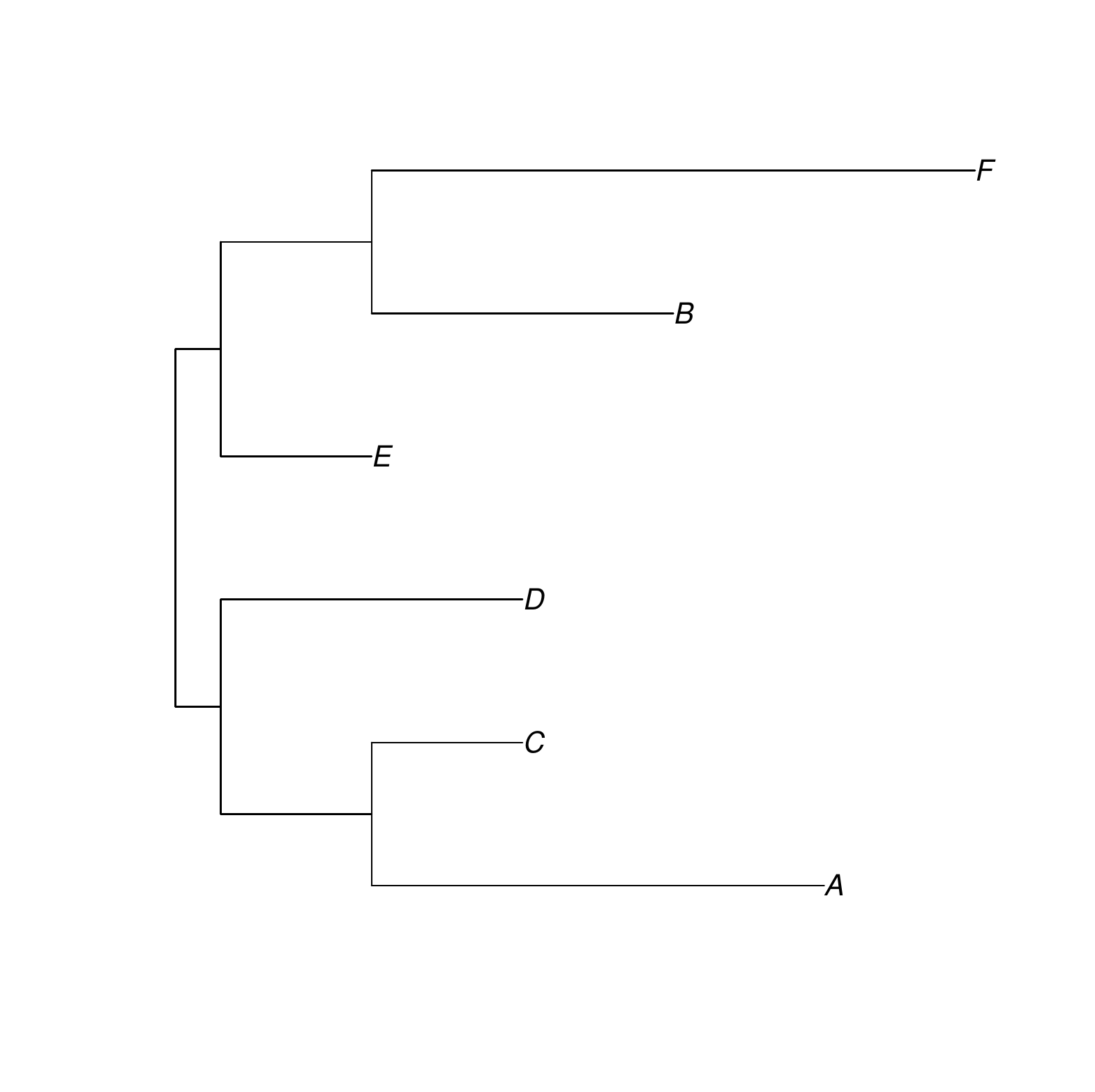}
\quad
\includegraphics[scale=0.4]{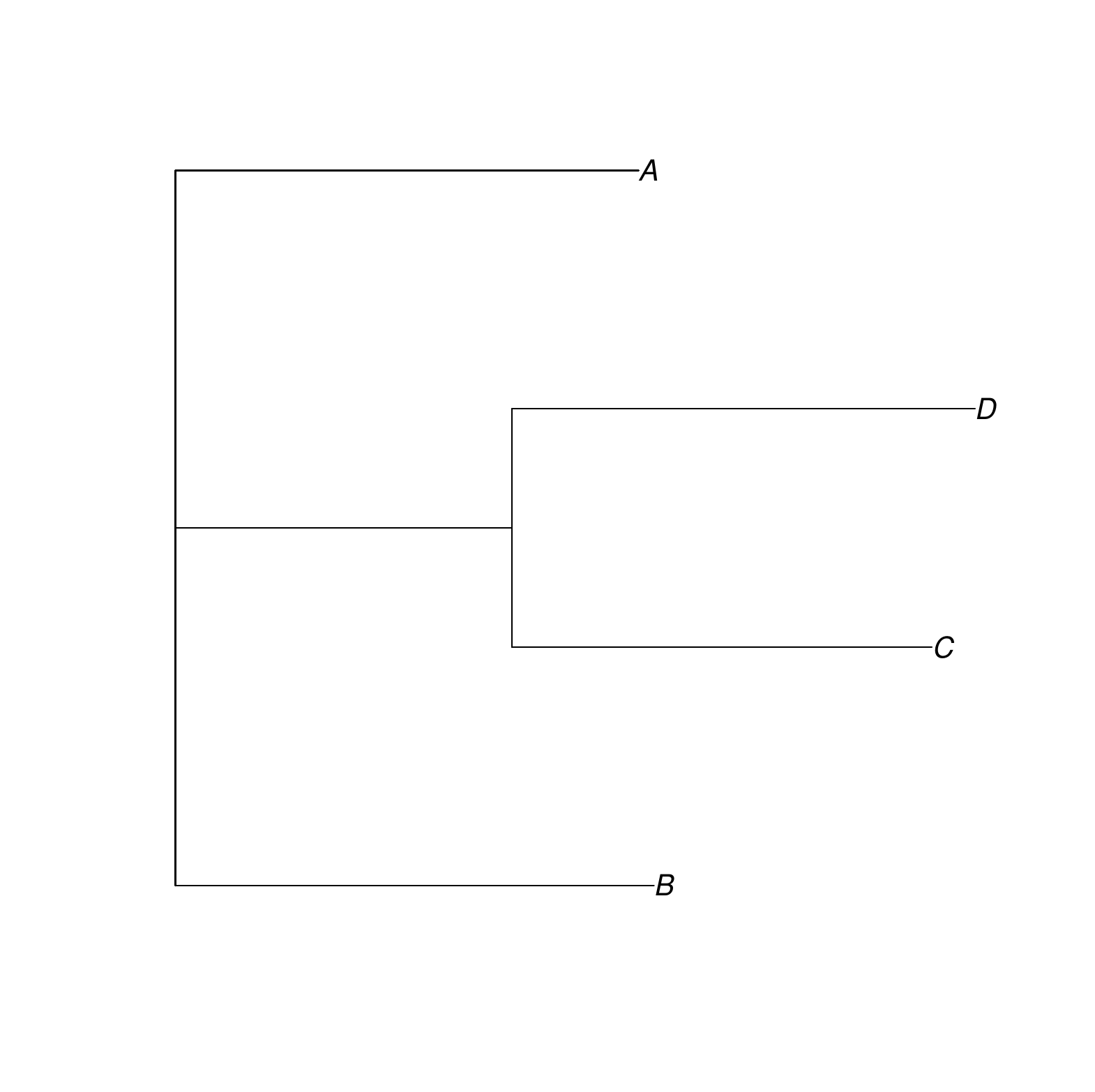}
\caption{\em True phylogenetic trees 
of the two synthetic data used as benchmarks: 
\textsf{Hadamard 1}  (left Panel) 
\textsf{Hadamard 2} (right Panel).} 
\label{Fig.2}
\end{figure}
\end{center}

We use as a first benchmark the 
synthetic data set \textsf{Hadamard 1} 
already employed in \cite{Felsenstein:2004}.
It consists of a sequence 1000 of amino acid
alignments of  six species, A, B, C, D, E and F simulated from a
$GTR + \Gamma$ model.  The true tree 
is
shown in the left Panel
of Figure \ref{Fig.2}.  
$GTR + \Gamma$ model 
is implemented in \textsf{MrBayes} 
which 
uses 
the Metropolis Coupling algorithm 
\citep{Alte:2004}
and 
provides
as output 
the 
simulated Markov Chain
and some evaluations of Bayesian model evidence 
in terms of posterior AM and 
marginal likelihood via HM.

The simulated Markov chain can 
be
used as a sample from the posterior
distribution in order to make 
inference on the model parameters.
For this model the 
whole parameter
space consists of 18 parameters.
In order to reduce the
autocorrelation and improve the convergence of the 1100000
sampled values, 100000 
have 
been discarded with thinning rate equal
to 10. 
We have recycled this
MCMC output 
to estimate
the marginal likelihood of the  $GTR + \Gamma$ model
for the known true topology
through IDR method.
In Table~\ref{Tab3}
we list the corresponding values of the IDR estimator on the log scale ($\log \hat{c}_{IDR}$), 
the Relative Mean Square Error estimate 
($\widehat{RMSE}_{\hat{c}_{IDR}}$) as in Equation (\ref{RMSE_hat})
and the confidence interval 
$\widehat{CI}$ for different perturbation masses $k$. In order to take into account the autocorrelation
of the posterior simulated values, a correction has been applied to
$\widehat{RMSE}_{\hat{c}_{IDR}}$
replacing $n$ in 
\eqref{RMSE_hat}
with the effective sample size 
given by
\begin{equation}
{n}_{ESS} = n \times \frac{1}{1+2 \sum_{i=1}^{I} \hat{\rho}_i}
\label{effss}
\end{equation}

Since the 
optimal corrected error
$\widehat{RMSE^{*}}_{\hat{c}_{IDR}}$ corresponds to a perturbation value
$k^{opt}=10^{-7}$, the IDR estimator (on a logarithmic scale) for the $GTR + \Gamma$ model is $\log \hat{c}_{IDR}=-7258.200$.
\begin{center}
\textbf{Table 3 about here}
\end{center}

We compare the results of the IDR method with
those obtained with the  Harmonic Mean (HM)
and 
the posterior Arithmetic Mean (AM). 

For $\hat{c}_{HM}$ and $\hat{c}_{AM}$, relative errors have been estimated respectively as 
\begin{equation}
\widehat{RMSE}_{HM}={\widehat{c}_{HM}} \sqrt{\widehat{Var}\left(\frac{1}{g(\theta)}\right)}
\label{rmse-HM}
\end{equation}

\begin{equation}
\widehat{RMSE}_{AM}=\frac{1}{\widehat{c}_{AM}}\sqrt{\frac{\widehat{Var}(g(\theta))}{n}}
\label{rmse-AM}
\end{equation}

Similarly to $\widehat{RMSE^{*}}_{\hat{c}_{IDR}}$
we have denoted with $\widehat{RMSE^{*}}_{\hat{c}_{HM}}$ and $\widehat{RMSE^{*}}_{\hat{c}_{AM}}$ 
the 
estimates of the
relative errors adjusted
with the effective sample size $n_{ESS}$.

The three methods produce somewhat different quantities 
although sometimes 
compatible once accounted for the estimation error. 
For each method, the Monte
Carlo relative error of the estimate has been computed 
re-estimating the
model $10$ times ($\widehat{RMSE}_{\hat{c},MC}$)
and recording the corresponding different values of 
$\hat{c}$. \\
Although it is known that under critical conditions 
such MC error 
is not sufficient to guarantee 
its precision 
it still remains a necessary premise 
for an accurate estimate. 
We have also looked at 
another precision measurement $\widehat{RMSE}_{\hat{c},Boot}$ based on bootstrap replications.
$\widehat{RMSE}_{\hat{c},Boot}$ is defined as
$$\widehat{RMSE}_{\hat{c},Boot}=\sqrt{\frac{1}{B}\sum_{b=1}^{B}\left(\frac{\hat{c}_{b}}{\hat{c}}-1\right)^{2}}$$
where $\hat{c}_{b}$ denotes the bootstrap replicate of the 
generic marginal likelihood estimate 
with one of the three alternative formulae
$\hat{c}_{AM}$, $\hat{c}_{HM}$ or $\hat{c}_{IDR}$ and $B=1000$.
 In order to account for the autocorrelation,
also $\widehat{RMSE}_{\hat{c},Boot}$ has been corrected using the effective sample size in (\ref{effss}).

Table~\ref{Tab4}  shows the obtained results.
 


\begin{center}
\textbf{Table 4 about here}
\end{center}

Indeed 
the estimates of the model evidence as well as 
the corresponding estimates of their relative errors
are very different. In fact, the
smallest 
relative 
error is obtained 
with the arithmetic mean of
the 
likelihood values
simulated from the posterior. 
We have already 
pointed out 
the fact that this posterior AM 
does not 
really 
aim at estimating the marginal likelihood, but we 
have nonetheless considered it in 
our examples
to verify 
how distant 
the corresponding values are and how different the conclusions 
and their strengths
can be when comparing alternative models via the 
posterior AM estimator.  
For the Harmonic Mean method
the estimated MC relative error is 
formidably high and unstable 
resulting in a serious warning on 
the reliability 
of the HM estimator
in this context.
On the other hand, the Inflated
Density Ratio approach seems to be a good compromise in terms of
order of magnitude of the 
error of the estimate $\hat{c}_{IDR}$
and robustness of its relative 
error estimation ranging from 0.15 with 
independent Monte Carlo re-estimation 
to  0.30 of the $\widehat{RMSE^{*}}_{\hat{c}}$. \\

\subsection{Hadamard 2: Bayes Factor computation}

We have also considered 
the \textsf{Hadamard 2} data
in \citep{Felsenstein:2004}
as a second benchmark synthetic data set. 
It
consists of 200 amino acids and four species, A, B, C, D.
This dataset 
have been simulated from the Jukes-Cantor model (JC69) and the true tree is
shown in the right Panel of Figure \ref{Fig.2}.\\
For the true
topology, we compute the
marginal likelihood for the JC69 model  and
for the GTR$+ \Gamma$ model: parameters lie in a 5-dimensional space
for the JC69 model and in 14-dimensional space for the GTR$+ \Gamma$ model.
The simulated values from the
Metropolis-Coupled algorithm have been rearranged to 
evaluate the model evidence of 
both
models using 
IDR, 
HM and the AM approach. As for the
\textsf{Hadamard 1} data, Monte Carlo RMSE have been 
also computed 
by
repeating the estimates 10 times. 
Table~\ref{Tab6} shows the results obtained respectively for
the \textbf{GTR}$+\Gamma$ and the \textbf{JC69} models.

\begin{center}
\textbf{Table 5 about here}
\end{center}

All methods 
produce  somewhat different 
results in terms 
of model evidence;
as for the previous example, the 
smallest 
relative 
Monte Carlo
error is associated 
once again with the Arithmetic Mean method. Also in this case,
estimates of the relative
errors of the Harmonic Mean method 
are always
larger than those produced by the
Inflated Density Ratio method. The corresponding Bayes Factors (on
logarithmic scale) for JC69 and GTR$+\Gamma$ are shown in Table~\ref{Tab8}.
Considering the reference values for the Bayes Factor defined in
\cite{Kass:1995}, all methods 
consistently give support to 
the Jukes-Cantor
model, which is known to be the true model. 
However,
we highlight that  
the strongest evidence in favor of the correct model 
corresponds to the Bayes Factor as estimated by the
Inflated Density Ratio .

\begin{center}
\textbf{Table 6 about here}
\end{center}

\subsection{Hadamard 2: tree selection}

\label{IDR-Tree-Section}

We now show how it is possible to extend the 
IDR approach for dealing with 
selecting competing trees when 
the topology is not fixed in advance. 
For a fixed substitution
model, competing trees can be compared by considering the evidence
of the data for a fixed tree topology.
The evidence in support of 
each tree
topology $\tau_{i} \in N_T$ can be evaluated in terms of its posterior
probability $p(\tau_{i}|X)$ derived from the Bayes theorem as
$$p(\tau_{i}|X) 
=
\frac{p(X|\tau_{i})\pi(\tau_{i})}{
\sum_{\tau \in N_T} p(X|\tau)\pi(\tau)}$$
where the experimental evidence 
in favor of the model $M_{\tau_{i}}$ 
with fixed tree topology $\tau_i$ 
is contained in the marginal likelihood
$$m(X|M_{\tau_{i}})= p(X|\tau_{i})=
\int_{\Omega_{\tau_i}} p(X|\omega_i,M_{\tau_{i}})\pi(\omega_i|\tau_{i})d\omega_i$$
where the continuous 
parameters $\omega_i \in \Omega_{\tau_i}$ of the evolutionary
process corresponding to $M_{\tau_{i}}$ 
are integrated out as nuisance parameters. 
Indeed 
when prior beliefs on trees are set equal $\pi(\tau_{i})=\pi(\tau_{j})$ 
comparative evidence discriminating 
$\tau_{i}$ against $\tau_{j}$, 
is summarized in the Bayes Factor 
\begin{equation}
\centering BF_{ij}=\frac{m(X|M_{\tau_{i}})}{m(X|M_{\tau_{j}})}.
\label{BF-Tree}
\end{equation}

We have considered the problem of selecting competing trees 
of a substitution model for \textsf{Hadamard 2} data.
In the previous Subsection,
we have verified the feasibility of
the Inflated Density Ratio approach in comparing
JC69 with the GTR$+ \Gamma$ model. Under a fixed topology
JC69 was favored.
Indeed we know that Hadamard 2 data was simulated from JC69 model 
with true topology labeled as $\tau=1$. 
Now we aim at comparing $N_{T}=3$ alternative topologies
within the JC69 model and we  
compute and compare the corresponding marginal likelihoods.
Results are shown in Table~\ref{Tab9}.

\begin{center}
\textbf{Table 7 about here}
\end{center}

Also in this case, the 
IDR 
exhibits 
the most convincing performance in terms of 
evidence in support of the true tree as well as 
precision and robustness 
of estimates.

\section{Discussion}
\label{sec:07}

In this paper, we 
investigate the possibility of using simple 
effective recipes for  
evaluating model evidence of competing models
of 
complex 
phylogenetic models. 
In a Bayesian framework,
several methods have been proposed in order to 
approximate
the marginal
likelihood of a single model 
and then eventually 
estimate 
the Bayes Factor
of two competing models. 

Probably, the most widely used methods 
to date are the thermodynamic
integration and the harmonic mean approach. The
thermodynamic integration has been proved to 
produce more reliable estimates 
of Bayes Factors of competing
phylogenetic models in a large varieties of 
contexts. 
Although this method
has the advantage of general applicability, 
it is computationally demanding  
and may require 
fine tunings and adjustments. 
Indeed, 
the simplicity of implementation
combined with a relatively 
light
computational burden 
are two appealing features which 
explain why the HM is still currently one of the most favorite 
option for routine implementation 
(see \cite{Reumont:2009}).
However, the simplicity of HM is often not matched with its accuracy 
and recent literature is highlighting 
unreliability of HM estimators 
in phylogenetic models 
\citep{Lartillot:2006}
as well as in more general 
biological applications
\citep{giro:2009}.
In this paper, we have provided evidence of 
improved effectiveness of a simple alternative
marginal likelihood estimator, 
the Inflated Density Ratio estimator (IDR), 
belonging to the class of 
generalized harmonic mean estimators.
It shares the original simplicity
and computation feasibility of the HM estimator but, 
unlike it, it 
enjoys
important theoretical properties, such as the finiteness of the variance.
Moreover it allows 
one to recycle 
posterior simulations 
which is particularly appealing in those contexts 
-- such as phylogenetic models --
where the computational burden of the simulation 
is heavier than the evaluation of the likelihood,
posterior densities and the like. 
Like all importance sampling techniques based on a single stream 
of simulations the computational burden
can be shared in a parallel computing environment 
reducing the computing time. Also 
the grid search for optimizing the estimated RMSE 
can be speeded up with a parallel 
evaluation for each inflated density.


We have verified  
the effectiveness 
of the IDR estimator 
in some of the most common phylogenetic 
substitution models
under different model complexity 
including mixed parameter space
and evaluated 
the comparative performance 
with respect to 
HM and posterior AM estimators.
In all circumstances 
the IDR estimator outperformed 
the
HM 
estimator in terms of precision and 
robustness of the estimates
and it is then an 
interesting candidate 
to be included in standard software as 
a simple and more reliable
model evidence output. 
Its simple implementation 
makes the IDR estimator a useful tool to be possibly 
used as 
a simple confirmation/benchmark  
even in those models 
where fine-tuned approximation tools 
such as thermodynamic integration 
are available and, when appropriately fine-tuned,
may
yield more precise estimates.

\newpage

\begin{table}[htbp]
\begin{center}
\begin{tabular}{ll}
\hline
Software & Marginal likelihood estimation method\\
\hline
BayesTraits & Harmonic Mean\\
BEAST & Harmonic Mean or bootstrapped Harmonic Mean \\
MrBayes & Harmonic Mean \\
PHASE & Harmonic Mean and Reversible Jump\\
PhyloBayes & Thermodynamic Integration under normal approximation \\
\hline
\end{tabular}
\end{center}
\caption{\em 
Availability of marginal likelihood estimation methods 
in Bayesian phylogenetics software: 
in spite 
of its inaccuracy, the harmonic mean estimator is still 
one of the most diffuse marginal likelihood estimation tool.}
\label{Tab1}
\end{table}

\newpage

\begin{small}
\begin{table}[htbp]
\centering
\begin{tabular}{lrrrr}
\hline
Distribution & $\log c$ & $\log \hat{c}_{IDR}$ & $\widehat{RMSE}_{\hat{c}_{IDR}}$ & $k^{opt}$ \\
\hline
& & & &\\ 
$N(\mu,\sigma)$ & {{{0}}} & {{0}} & $10^{-4}$ & $10^{-4}$ \\
$N_{100}(\mu,\sigma)$ & {{{0}}} & {{0}} & 0.01 & $10^{23}$ \\
& & & &\\
$SN(\mu,\sigma,\tau)$ & {{{0}}} & {{$10^{-4}$}} & $0.004$ & $10^{-4}$ \\
$SN_{5}(\mu,\sigma,\tau)$ & {{{3,467}}} & {{3.444}} & $0.014$ & $20$ \\
$SN_{30}(\mu,\sigma,\tau)$ & {{{2,302}}} & {{2.403}} & $0.047$  & $10^{4}$  \\
& & & &\\ 
Mix $N_{2}$ & {{{2.079}}} & {{{2.078}}} & 0.003 & 0.01 \\
Mix $N_{3}$ & {{{2.772}}} & {{{2.766}}} & 0.002 & 1 \\
Mix $N_{10}$ & {{{0}}} & {{{0.056}}} & 0.012 & 2 \\
\hline
\end{tabular}
\caption{\em Performance 
of IDR approach with $n=10^5$ i.i.d. draws simulated from distributions 
with known normalizing constants:  
univariate and multivariate gaussian
distributions (up to 100 dimension),
univariate and multivariate skew normal distributions 
($SN$) (30 dimension) 
and multivariate mixtures of two Normal components (Mix $N$)
(in dimensions 2,3 and 10).
The value $\log c$ represents the logarithm 
of the true value of the normalizing constant; 
$\log \hat{c}_{IDR}$ is the value of the estimated
normalizing constant on a logarithmic scale. 
$k^{opt}$ is the optimal inflation
coefficient which minimizes the Relative Mean Square Error 
$\widehat{RMSE}_{\hat{c}_{IDR}}$
computed as in (\ref{RMSE_hat}). 
For the SN case a small 
sensitivity study (not reported here) showed that the performance of the method 
is robust to different $(\mu,\sigma,\tau)$ parameter choices.}
\label{Tab2}
\end{table}
\end{small}

\newpage

\begin{table}[htbp]
\begin{center}
\begin{tabular}{cccccccc}
  \hline
& $k$ & $\log \hat{c}_{IDR}$ &$\widehat{RMSE}_{\hat{c}_{IDR}}$  &$\widehat{RMSE^{*}}_{\hat{c}_{IDR}}$ & $\widehat{CI}$ \\
 \hline
&$10^{-10}$ & -7264.438 & 0.1710 & 0.4515 &  $[-7264.852, -7263.718]$\\
&$10^{-9}$ & -7262.150  & 0.1689 & 0.4514 & $[-7262.560, -7261.443]$ \\
&$10^{-8}$ & -7259.939  & 0.1602 & 0.3664 & $[-7260.332, -7259.284]$ \\
($*$)&$10^{-7}$ & -7258.200  & 0.1178 & 0.3008& $[-7258.503, -7257.764]$ \\
&$10^{-6}$ & -7257.554  & 0.1407 & 0.3694&  $[-7257.907, -7257.006]$ \\
\hline
\end{tabular}
\end{center}
\caption{\em Inflated Density Ratio method 
applied to \textsf{Hadamard 1}
data with a GTR+$\Gamma$ model 
with a 18-dimensional parameter space. 
IDR estimates on the log scale
for a small regular grid of perturbation values. 
The 
relative mean square errors $\widehat{RMSE}_{\hat{c}_{IDR}}$ 
are computed as in (\ref{RMSE_hat}) 
without accounting for autocorrelation.
$\widehat{RMSE^{*}}_{\hat{c}_{IDR}}$
are the relative mean square errors 
corrected for the autocorrelation. 
$\widehat{CI}$ are  
confidence intervals on a log scale. 
Since the smallest error in the grid corresponds to a perturbation value
$k^{opt}=10^{-7}$, the IDR estimator 
for the $GTR + \Gamma$ model is taken to be $\log \hat{c}_{IDR}=-7258.200$. 
} 
\label{Tab3}
\end{table}

\newpage

\begin{table}[htbp]
\begin{center}
\begin{tabular}{cccccc}
  \hline
Method & $log(\widehat{c})$ & $\widehat{RMSE}_{\hat{c}}$ & $\widehat{RMSE}_{\hat{c},MC}$ & $\widehat{RMSE}_{\hat{c},Boot}$ & $\widehat{RMSE^{*}}_{\hat{c}}$ \\
 \hline
IDR & -7258.200 & 0.1178 & 0.1538  &  0.1698 & 0.3008 \\
HM & -8365.509 &  173.2080 & $>10^{10}$ & $>10^{100}$ & 296.3475  \\
AM  & -7204.245 &  0.0197 & 0.0119  & 0.0202 & 0.065 \\
\hline
\end{tabular}
\end{center}
\caption{\em \textsf{Hadamard 1} data: marginal likelihood estimates
obtained with the Inflated Density Ratio method $\widehat{c_{IDR}}$,
with the Harmonic Mean approach $\widehat{c_{HM}}$ and with the
Arithmetic Mean approach $\widehat{c_{AM}}$. The estimates
are based on $n=10^6$ from $10^7$ simulated values. Three different 
RMSE estimates are provided:
$\widehat{RMSE}_{\hat{c}}$ has been computed as in (\ref{RMSE_hat}) for IDR, (\ref{rmse-AM}) for AM and (\ref{rmse-HM}) for HM; $\widehat{RMSE}_{\hat{c},MC}$ comes from 10 Monte
Carlo independent replicates of the estimation; $\widehat{RMSE}_{\hat{c},Boot}$
is a boostrap approximation of RMSE (B=1000);
$\widehat{RMSE^{*}}_{\hat{c}}$  
is the estimated RMSE corrected for the autocorrelation as in (\ref{effss}). }
\label{Tab4}
\end{table}

\newpage

\begin{table}[htbp]
\begin{center}
\begin{tabular}{cccccc}
  \hline
  \multicolumn{6}{|c|}{GTR$+\Gamma$}\\
\hline \\
  Method & $log(\widehat{c})$ & $\widehat{RMSE}_{\hat{c}}$ & $\widehat{RMSE}_{\hat{c},MC}$ & $\widehat{RMSE}_{\hat{c},Boot}$ & $\widehat{RMSE^{*}}_{\hat{c}}$ \\
  \hline
  IDR &  -611.8571 & 0.1153 & 0.1087 & 0.1175  & 0.3608    \\

HM & -594.648 & 31.5330 & 0.1329 &  0.3488 & 141.3285    \\

AM & -588.286 & 0.0184 & 0.0863 & 0.0187 & 0.0826   \\
\hline
  \multicolumn{6}{|c|}{JC69}\\
\hline\\
Method & $log(\widehat{c})$ & $\widehat{RMSE}_{\widehat{c}}$ & $\widehat{RMSE}_{\widehat{c},MC}$ & $\widehat{RMSE}_{\widehat{c},Boot}$ & $\widehat{RMSE^{*}}_{\widehat{c}}$ 
\\
 \hline
IDR &  -595.5919 &  0.0068 & 0.0161 & 0.0081  & 0.0179     \\
HM & -589.0289   & 34.6759  & 0.1415 & 0.6787 & 59.5918   \\
AM & -589.4194   &  0.0057 & 0.0146 &  0.0056 & 0.0182   \\
\hline
\end{tabular}
\end{center}
\caption{\em \textsf{Hadamard 2} data: marginal likelihood estimates
of the GTR$+\Gamma$ model ($ \omega = \Re^{14}$) obtained with the IDR method $\widehat{c_{IDR}}$,
with the HM approach $\widehat{c_{HM}}$ and with the
AM approach $\widehat{c_{AM}}$.  The estimates
are based on $n=10^6$ from $10^7$ simulated values. Three different 
RMSE estimates are provided:
$\widehat{RMSE}_{\hat{c}}$ has been computed as in (\ref{RMSE_hat}) for IDR, (\ref{rmse-AM}) for AM and (\ref{rmse-HM}) for HM; 
$\widehat{RMSE}_{\hat{c},MC}$ comes from 10 Monte
Carlo independent replicates of the estimation; $\widehat{RMSE}_{\hat{c},Boot}$ is a bootstrap approximation
of RMSE (B=1000);
$\widehat{RMSE^{*}}_{\hat{c}}$ 
is the RMSE corrected for the autocorrelation as in (\ref{effss}).}
\label{Tab6}
\end{table}

\newpage

\begin{table}[htbp]
\begin{center}
\begin{tabular}{ccccc}
  \hline
Method & $log(\widehat{BF}_{GTR+\Gamma-JC69})$ & $\widehat{CI}^{MC}_{log(\widehat{BF})}$\\
 \hline
IDR & 16.2652 & $[16.1726,16.3578]$\\
HM & 5.6241 & $[5.0206, 5.1546]$\\
AM & 1.1334 & $[1.0617, 1.2051]$\\
  \hline
\end{tabular}
\end{center}
\caption{\em \textsf{Hadamard 2} data: Bayes Factors computed with
IDR, HM and AM approach.  The estimates
are based on $n=10^6$ from $10^7$ simulated values. $\widehat{CI}^{MC}_{log(\widehat{BF})}$ is the confidence interval 
obtained as $log(\hat{BF})\pm 2\cdot SD_{MC}(log(\hat{BF}))$. $log(\hat{BF})$ is obtained
by averaging the estimated Bayes Factors (on logarithmic scale) of $10 \times 10$ possible pairings of 10 MC replicates.
$SD_{MC}(log(\hat{BF}))$ is computed as the standard error of the estimated Bayes Factors (on logarithmic scale) in $10 \times 10$ possible combinations of  10 MC replicates.}
\label{Tab8}
\end{table}

\newpage

\begin{table}[htbp]
\begin{center}
\begin{tabular}{ccccc}
  \hline
$\tau$ & $Size$ & $\widehat{CI}^{MC}_{log(\widehat{BF}_{IDR})}$& $\widehat{CI}^{MC}_{log(\widehat{BF}_{HM})}$) & $\widehat{CI}^{MC}_{log(\widehat{BF}_{AM})}$\\
 \hline
  $log(BF_{12})$ & $10^4$ & $[3.511,3.599] $ & $[2.37,4.066]$& $[2.929,2.989]$\\
$log(BF_{13})$ & $10^4$ & $[3.817,3.901]$ & $[2.503,3.163]$& $[2.053,3.131]$\\
  \hline
\end{tabular}
\end{center}
\caption{\em \textsf{Hadamard 2}
data: the Bayes Factor is computed in order to compare competing
topologies 
The Bayes Factor is approximated with
the IDR method ($BF_{IDR}$), the HM  
($BF_{HM}$) and the AM ($BF_{AM}$) approach.
 $\widehat{CI}^{MC}_{log(\widehat{BF})}$ is the confidence interval 
obtained as $log(\hat{BF})\pm 2\cdot SD_{MC}(log(\hat{BF}))$. $log(\hat{BF})$ is obtained
by averaging the estimated Bayes Factors (on logarithmic scale) of $10 \times 10$ possible combinations of 10 MC replicates.
$SD_{MC}(log(\hat{BF}))$ is computed as the standard error of the estimated Bayes Factors (on logarithmic scale) in $10 \times 10$ possible combinations of  10 MC replicates.}
\label{Tab9}
\end{table}

\end{document}